\documentstyle[manuscript,aps]{revtex}
\newcommand{\beq}{\begin{equation}}
\newcommand{\eeq}{\end{equation}}
\newcommand{\beqa}{\begin{eqnarray}}
\newcommand{\eeqa}{\end{eqnarray}}
\newcommand{\beqar}{\begin{eqnarray*}}
\newcommand{\eeqar}{\end{eqnarray*}}

\def \A {{\bf A}}
\def \B {{\bf B}}
\def \T {{\bf T_A}}
\def \la {\langle}
\def \ra {\rangle}

\def \x {{\bf x}}

\def \p {{\bf p}}

\def \Q {{\bf Q}}
\def \P {{\bf P}}
\def \e {\epsilon}
\def \H {{\bf H}}
\def \X {{\bf X}}
\def \proj {{\bf \Pi}}
\begin{document}
\input epsf

\title{ \bf\Large
Minimum Inaccuracy for Traversal-Time
}
\author{ { 
J. Oppenheim$^{(a)}$,\footnote{\it jono@physics.ubc.ca}
B. Reznik$^{(b)}$,\footnote{\it reznik@t6-serv.lanl.gov}
and W. G. Unruh$^{(a)}$}\footnote{\it unruh@physics.ubc.ca \\
}  
{\ } \\
(a) {\it \small   Department of Physics,  University of British
Columbia,
6224 Agricultural Rd. Vancouver, B.C., Canada
V6T1Z1}\\
(b) {\it \small Theoretical Division, T-6, MS B288, 
Los Alamos National Laboratory, Los Alamos, NM, 87545}
}

\maketitle

\begin{abstract}

{

Using various model clocks it has been shown that the time-of-arrival
cannot be measured more accurately than $\delta T_A>1/E_p$ where $E_p$ is
the kinetic energy of a free particle.  However, this result has never
been proved.  In this paper, we show that a violation of the above
limitation for the traversal-time implies a violation of the Heisenberg
uncertainty relation.  

} 

\end{abstract}

\newpage

\section{Introduction}

In \cite{aharonov}, we considered various clock models for measuring
the time it takes for a free particle to arrive to a given location $x_A$.
Because the energy of the clock increases with its precision, we argued
that 
the accuracy of a time-of-arrival
detector cannot be greater than $1/E_p$, where $E_p$ is the kinetic energy
of the particle.
Measurements of traversal-time \cite{peres} are analogous to that of 
time-of-arrival.  One tries to measure how long it takes a particle
to travel between two fixed locations $x_1$ and $x_2$.  Although 
no proof 
has yet been found for the restriction on time-of-arrival accuracy, in 
this paper we are able to show that a necessary minimum inaccuracy on 
traversal-time measurements is given by
\beq
\delta T_F > 1/E_p.\label{eq:uncertainty}
\eeq 
We do this by arguing that a traversal-time measurement is 
also a simultaneous measurement of position and momentum, and that 
(\ref{eq:uncertainty}) is required in order to preserve the Heisenberg
uncertainty relationship.  Note however that (\ref{eq:uncertainty}) is
not analogous to the Heisenberg Energy-time uncertainty relationship.
It reflects the inherent inaccuracy of every individual measurement,
while the Heisenberg uncertainty relationships refer to well-defined
and perfectly accurate measurements
made on ensembles.

The article proceeds as follows. In section II we motivate the notion that 
traversal-time is a measurement of momentum by looking at measuring 
the traversal-distance.  In section III
we discuss a physical model for measuring the traversal-time, 
and show the relation between (\ref{eq:uncertainty}) and the uncertainty principle.
The main result of this paper is given 
in Section IV, where we provide a model independent derivation of 
(\ref{eq:uncertainty}),
as well as a qualitative proof. 
\section{Measuring momentum through traversal-distance}

The measurement of  traversal-distance may be considered 
the space-time  ``dual'' of  the measurement of traversal-time: instead of 
fixing $x_1$ and $x_2$ and measuring $t_F= t_2-t_1$, one
fixes $t_1$ and $t_2$ and measures $x_F= x_2-x_1$.
It is  instructive to examine first this  simpler case of 
traversal-distance and point out the similarities and the 
differences.

Unlike the case of traversal-time, a measurement of traversal-distance
can be described by the standard von Neumann interaction.
For a free particle the Hamiltonian is 
\beq
\H = \frac{\p^2}{2m} + \Q \x \biggl[\delta(t-t_1)-\delta(t-t_2) \biggr]
\eeq
where $\Q$ is the coordinate conjugate to the pointer variable
$\P$. The change in $\P$ yields the traversal-distance:
\beq
\P(t>t_2) - \P_0 = \x(t_2) - \x(t_1)= \x_F.
\eeq
 
However  the measurement of the traversal-distance provides
additional information: it also determines the momentum $\p$
of the particle $during$ the time interval $t_1<t<t_2$.
From the equations of motion we get:
\beq
\p(t) = 
\left\{
\begin{array}{ll}
	\p_o, & \mbox{$t<t_1$ or $t >  t_2$} \\ 
	\p_o -\Q,  & \mbox{$t_1 < t <t_2$}
\end{array}
\right.
\label{momentum}
\eeq
and 
\beq
\x(t) = 
\left\{
\begin{array}{ll} 
x_0 + 	{\p_o\over m}t,  & \mbox{$t \leq t_1$ } \\ 
x_0 + {\p_0\over m}t_1 + {\P_0-\Q\over m}(t-t_1),  & \mbox{$t_1 \leq t \leq t_2$}
\end{array}
\right.
\eeq
and therefore,
\beq
m{\P(t>t_2) - \P_0\over t_2-t_1 } = \p_0 - \Q = \p(t_1\leq t\leq t_2).
\eeq
Thus, one can determine simultaneously and to arbitrary accuracy
the traversal-distance and the momentum in  intermediate times.
This, of course, does not contradict the uncertainty relations, because 
$\p$ commutes with $\x_F$, and $\x$ remains completely uncertain.
Similarly, in the case of the traversal-time we shall see that the
measurement determines also the intermediate time  momentum, 
however unlike the present case, since the particle has to be
in the interval $x_2-x_1$ during the traversal, 
it is also a measurement of the location. This indicates that, 
in the latter case, in order not to violate the uncertainty
principle, the accuracy with which $T_F$ or $p$ may
be measured must be limited.

\section{Measuring traversal-time}
	In quantum mechanics, classical observables such as position, 
momentum and energy are represented by an operator $\A$  whose eigenvalues
give the possible outcomes of a measurement.  However, some classical
observables, such as time \cite{pauli} and time-of-arrival
\cite{aharonov}\cite{allcock} cannot be represented by operators. 
For example, for time-of-arrival, one can use the correspondence 
principle to find the operator (up to ordering difficulties) 
\beq
\T=m(\frac{1}{\p}\x+\x\frac{1}{\p}).
\eeq
However it turns out that due to the singularity at $p=0$,
the eigenstates of this operator are not orthogonal
and therefore $\T$ is not Hermitian.
One could regularize this operator in some way \cite{Rovelli}
however the resulting operator is unphysical. 
Measuring this  operator is not
equivalent to physically measuring the time-of-arrival \cite{aharonov}.

For traversal-time the situation
is similar.  The classical equations of motion suggest
that a traversal-time operator might be given by
\beq
{\bf T_F}=\frac{mL}{\p},
\eeq
where $L=x_2-x_1$.
Like the time-of-arrival operator, this operator is undefined at $p=0$,
and again  different outcomes are found in a direct
measurement of $T_F$ and a measurement of a regularized ${\bf T_F}$.  
One can measure the momentum at any time, so there is
no reason to believe that the particle actually travelled between the
two points in the time $t_F$.  A measurement of $1/\p$ will result in the 
particle's position being spread over all space, so there is no finite
amount of time one could wait before being certain that the particle
went between the two fixed points.  
For example, after the measurement of
$1/\p$, the potential between $x_1$ and $x_2$ might change.
General traversal-time 
operators would require that one knows the Hamiltonian not only in the 
past, but also in the future.  If one measures the traversal-time
operator above, then one has to have faith that the Hamiltonian
will not change after the time of the measurement $t_o$ to
$t\rightarrow\infty$.

It is also commonly accepted that the dwell time operator \cite{dwell}, 
given by
\beq
{\bf \tau_D} = \int_0^\infty dt \proj(t)
\eeq
where
\beq
\proj(0)= \int_{x_1}^{x_2} |x\ra\la x|  
\eeq
can be used to compute the traversal time\footnote{in our case, where there
is no potential barrier, the dwell time, and traversal time are equivalent}.  Such a quantity however,
cannot be measured, since the operator $\proj(t)$ does not commute with
itself at different times\cite{tmeas}
\beq
[\proj(t),\proj(t')] \neq 0.
\eeq
Therefore, one must measure the traversal-time in a more physical way. One
must demand that if we measure the traversal-time to be $t_F$, then 
the particle must actually traverse the distance between $x_1$ and $x_2$ 
in the time given by the traversal-time measurement.  For example, one could
have a clock which runs when the particle
is between $x_1$ and $x_2$ given by the Hamiltonian
\cite{peres}\cite{unruh}
\beq
\H=\frac{\p^2}{2m}+V(\x) \Q
\eeq
where the traversal-time is given by the variable $\P$ conjugate to 
$\Q$ 
and the potential $V$ is equal to $1$ when 
$x_1\leq x \leq x_2$ and $0$ everywhere else
\footnote{The Hamiltonian for this ideal clock is unbounded from below. 
To remedy this, once could consider a Larmor clock with a 
bounded Hamiltonian $H_{clock}=\omega {\bf J_z}$ \cite{peres}. When the
particle enters the magnetic field, its spin
precesses in the zy-plane. The clock's resolution can be made arbitrarily 
fine by increasing $J_z$.}
.
In the Heisenberg picture, the equations of motion are
\beq
\dot{\x}=\p/m,\,\,\,\,\dot{\p}=-\Q(\delta(\x-x_1)-\delta(\x-x_2))
\eeq\beq
\dot{\P}=V(\x), \,\,\,\, \dot{\Q}=0.
\eeq
The particle's momentum is disturbed during the measurement.
\beq
\p ' = \sqrt{\p^2-2m\Q} \label{eq:pmeas}
\eeq  
where $\p '$ is the particle's momentum during the measurement, and
$\p$ is the undisturbed momentum.
However if the interaction is weak $Q\ll E_p$, then after a sufficient time, 
the clock will read the undisturbed traversal-time
\beqa
\P(t\rightarrow\infty)-\P(0) & \simeq & \int^\infty_0
V\left( \x(0)-\frac{\p_o t}{m}\right) dt \nonumber\\
&=& \frac{m(x_2-x_1)}{\p}
\eeqa
If we require an accurate measurement of the traversal-time,
then a small $dP$ will result in large values of the coupling $Q$.
If $Q$ is too large, the clock can 
reflect the particle at $x_1$ and one will obtain a traversal-time
equal to $0$.   
This therefore imposes a restriction on the accuracy
with which one can measure the traversal-time.  As in Ref. \cite{aharonov}
we find that 
\beq
\delta T_F > 1/E_p \label{eq:flightuncertainty}
\eeq
is required in order to be able to measure the traversal-time,
and 
\beq
\delta T_F \gg 1/E_p
\label{limit}
\eeq
in order to measure the undisturbed value of the traversal-time.

Let us show that the above conditions are consistent with the 
uncertainty relations for the position and momentum.
If (\ref{limit}) is satisfied,  we have $Q\ll E$, and by eq.
(\ref{eq:pmeas}) the momentum during the measurement is
\beq
\p ' \simeq \p - \frac{m}{\p} \Q.
\eeq
Thus  during the
measurement, the momentum will be uncertain  by an amount
\beq
dp'\simeq \frac{m}{p_o} dQ.
\eeq
In order to know whether the particle entered our detector, we need
to be able to distinguish between the case where the pointer is at
its initial position $P=0$, and the case where the particle has gone
through the detector $P=t_F=\frac{mL}{p_o}$.  
We therefore need the condition
\beq
dP<\frac{mL}{p_o}.
\eeq
Since at best we have $dP=1/dQ$, we find
\beq
dp'dx = dp' L > 1.
\eeq

The uncertainty relation (\ref{eq:flightuncertainty}) 
only applies to this particular model clock - 
it might be possible to accurately measure the 
traversal-time in some clever way.
In the following section we will show that this cannot be done, by 
demonstrating that this uncertainty applies to all measurements of 
traversal-time.  

Finally, we should note that a traversal-time 
detector could be made by measuring the time-of-arrival to $x_1$ and the
time-of-arrival to $x_2$. This would require two time-of-arrival
clocks each with its own inaccuracy, whereas the model above only
has one clock.       
%
%
%
\section{minimum uncertainty for traversal-time }
Before proceeding with the argument, we should be clear to distinguish
between different types of uncertainties.  For an operator $\A$,
there exists a kinematic uncertainty which we will denote by
$d\A$ given by
\beq
d\A = \la \A^2\ra - \la\A\ra^2.
\eeq
This is the uncertainty in the distribution of the observable $A$.
There is also the inherent inaccuracy in the measuring device.  This
is the relevant uncertainty in equations (\ref{eq:uncertainty}) and 
(\ref{eq:flightuncertainty}).  
It refers to the uncertainty in the initial state of the 
measuring device's pointer position $P$, and
we will denote it by $\delta A$.  For our measuring devices we have
\beq
\delta A = dP_o
\eeq
This uncertainty applies to each
individual measurement.  Lastly, there is the 
uncertainty $\Delta A$ which applies to the spread in measurements
made on the ensemble.  Given a set $A_M$ of experiments $i=1,2,3...$ which 
yield results $A_i$, we have
\beq
\Delta A  = <A_M^2>-<A_M>^2. 
\eeq
This uncertainty includes a component due to the 
kinematic uncertainty of the attribute of the system, and also
the inaccuracy of the device.  For our measuring device, the kinematic spread 
in the pointer position at the end of each experiment gives $\Delta A$
\beq
\Delta A = dP_f
\eeq

The Heisenberg uncertainty relationship $dA dB>1$ applies to 
measurements on ensembles.  Given an ensemble, we measure $\A$ on
half the ensemble and $\B$ on the other half.  The uncertainty 
relation also applies to simultaneous measurements. If we
measure $\A$ and $\B$ simultaneously on each system in the ensemble, 
then the distributions of $\A$ and $\B$ must still satisfy the 
uncertainty relationship.  

Returning now to the traversal-time, we see that it can be interpreted as
a simultaneous measurement of position and momentum. 
We know the particle's momentum $p$
during the time that it was between $x=x_1$ and $x=x_2$ from
the classical equations of motion
\beq
t_F=\frac{mL}{p}.\label{eq:classtime}
\eeq 
In other words, eigenstates of momentum must have traversal-times given
by equation (\ref{eq:classtime}). 
This measurement of momentum is analogous to the measurement
described in section II.  Instead of measuring the change in
position at two specified times $t_1$ and $t_2$, we are now
measuring the difference in arrival times after specifying two different
positions $x_1$ and $x_2$.  During
the measurement, we also know that particle is somewhere between
$x=x_1$, and $x=x_2$.  ie. we know that $x=\frac{x_1+x_2}{2} \pm L/2$.

The uncertainty relationship also applies to these measured 
quantities $\Delta x \Delta p > 1 $.  This essentially means that a
detector
of size $L$ will disturb the momentum by at least $2/L$, so that repeated
measurements on an ensemble will give $ \Delta p > 2/L$.  The position of the detector $\X$ computes with the momentum of the
particle $\p$ \cite{ab} however, we demand that the particle actually travel the
distance $L$.  The particle must actually be inside the detector during
the measurement.  As a result, $\X$ must be coupled to the position
$\x$ of the particle and so a measurement of $\X$ is also a measurement of 
$\x$. This is what we mean by a local interaction.  

We can see qualitatively, why we expect (\ref{eq:uncertainty}) to be true.
During the measurement of traversal time, the momentum will be disturbed
by an amount
\beq
dp>2/L.
\eeq
If this disturbance is small, then from (\ref{eq:classtime}) we expect this 
will cause an inaccuracy given by
\beqa
\delta T_F &=& \frac{mL}{p^2} dp \nonumber\\
&>& 1/E_p
\eeqa

We now proceed with the more rigorous argument.
We imagine a traversal-time detector which has an inaccuracy given by 
$\delta T_F$. Measurements can then be carried out on arbitrary ensembles
with arbitrary Hamiltonians. 
We will show that by choosing this ensemble appropriately, the 
uncertainty relationship $\Delta x \Delta p >1$ 
can be violated, unless the traversal-time obeys the  relationship
given by (\ref{eq:uncertainty}).

We assume that initially, the pointer on our traversal-time detector is
given by
\beq
P_o = \e
\eeq
where $\e$ is a small random number which arises because of the 
initial inaccuracy of the clock.  ie. the distribution
of $\e$ is such that $\la\e\ra=0$ and
the clock's initial inaccuracy in pointer position is $dP_o^2 = \la\e^2\ra$.
It is important to note that this inaccuracy is fixed as an
initial condition before any measurements are made.  It is a property of
the device, and does not depend on the nature of the ensemble upon which
we will be making measurements. For a free Hamiltonian, a measurement of the
traversal-time will result in
a final pointer position given by
\beq
P_f=P_o + \frac{mL}{p}  \label{eq:pointer}
\eeq
where $p$ is the momentum of the particle in the absence of any measuring
device.  For eigenstates of $\p$ (or states peaked highly in $p$), we demand
that the traversal-time be given by the 
classically expected value
\footnote{It is possible to include small deviations from
the classical value, by including an additional term in (\ref{eq:pointer}).
These fluctuations need to average to zero
in order to satisfy the correspondence principle.  For small fluctuations,
the following discussion is not altered. }
Recall that the kinematic spread in the particle's momentum
is given by $d\p = \la\p^2\ra-\la\p\ra^2$.  
A measurement of the traversal-time for a particular experiment $i$ can 
take on the values
\beqa
T_i &= &P_f \nonumber\\
&=&\frac{mL}{p} + \e 
\eeqa
%
A given measurement $T_i$ will allow us to 
infer the momentum of the particle $p_i$ during the measurement
%
%
\beq
p_i(T_i)=\frac{m L}{T_i}=\frac{m L p}{m L + p \e }.
\eeq
The average value of any power $\alpha$ of the measured momentum is 
\beq
\la p_M^\alpha \ra=\int \left( \frac{m L p}{m L + p \e}\right) ^\alpha
f(p) g(\e) dp d\e
\eeq
where $f(p)$ gives the distribution of the
particle's momentum and $g(\e)$ is the distribution of the fluctuations.
We now choose $m$ of the ensemble so that we always have 
\beq
\e  p\ll m L .\label{eq:condition}
\eeq
Indeed for the example given in the previous section,
for every given $\epsilon$ and $p$, we can increase $E_p$ by
choosing a sufficiently large $m$, and reach this limit.
This limit ensures that $\la p_M \ra$ never diverges,  
and simplifies our calculation by allowing us to write 
\beq
\la p_M^\alpha \ra \simeq\int \left( p-\frac{\e p^2}{m L} \right)^\alpha
f(p) g(\e) dp d\e
\eeq
For $\alpha=1$ we find
\beqa
\la p_M \ra &\simeq& \la p \ra-\frac{\la \e\ra\la p^2\ra}{m L} \nonumber\\
&=& \la p\ra.
\eeqa
 For $\alpha=2$ we find
\beqa
\la p_M^2 \ra & \simeq & \int \left(p^2-2\frac{\e p^3}{m L}+(\frac{\e p^2}{m L})^2\right)
f(p) g(\e) dp d\e \\
&=& \la p^2\ra + \frac{\la p^4\ra \la \e^2\ra  }{(m L)^2}.
\eeqa
This gives us
\beqa
\Delta p^2 &= &\la p_M^2 \ra-\la p_M \ra^2 \nonumber\\
&=& \frac{\la p^4\ra\delta T_F^2}{(mL)^2}+dp^2
\eeqa
Since
\beq
(dE)^2=\frac{\la p^4 \ra }{4m^2}-\la E \ra ^2
\eeq
we find
\beq
\Delta p ^2 = (\frac{2\delta T_F }{L})^2(dE^2+\la E \ra^2)
+ dp^2.
\eeq

Finally, we arrive at the relation 
\beq
(\Delta x \Delta p)^2=\delta T_F^2(\la E\ra^2+dE^2)+\frac{L^2}{4}dp^2.
\eeq
The uncertainty relation
\beq
\Delta x \Delta p > 1
\eeq
then  implies
\beq
\delta T_F ^2 > \frac{1-\frac{1}{4} L^2 dp^2}{\la E\ra^2+dE^2}.
\eeq
Now we note that we can arrange our experiment with 
$L dp$ arbitrarily small, by choosing $dp$ of the ensemble
arbitrarily small.  
As a result, in order
to ensure that Heisenberg's uncertainty relation is never
violated, we must have
\beq
\delta T_F > \frac{1}{\sqrt{\la E\ra ^2+dE^2}}. \label{eq:generaluncert}
\eeq
The condition (\ref{eq:condition}) and (\ref{eq:generaluncert}) 
imply that we have 
$dE\ll E$, so we can write
\beq
\delta T_F > \frac{1}{\la E \ra}.
\eeq

It is interesting to note that since the momentum operator 
commutes with the free Hamiltonian, the restriction on traversal-time
measurements only comes from the dynamical considerations given above.
\section{Conclusion}
We have seen that 
the measurement of the traversal-time given two positions cannot
be made arbitrarily accurate. This strongly suggests that 
the limitation on measurements of arrival times is a general rule
and not just an artifact of the types of models considered so far.
Operators for both these quantities are
singular or don't seem to correspond to physical (continuous) processes. 
The case of traversal-time is different from time-of-arrival in that
the semi-bounded spectrum of the Hamiltonian does not seem to play
an important role in the restriction on measurement accuracy.  
The accuracy restriction on traversal-time is particularly important
for experiments on barrier tunnelling time.  One usually uses
a physical clock to measure the time it takes for a particle to 
travel from one location to another, with a barrier situated somewhere
between the two locations \cite{landauer}\cite{unruh}. 
These measurements need to be inherently inaccurate, because if one tries to measure the tunnelling time too accurately, the particle will not
be able to tunnel. Our result concerning traversal-time indicates that the barrier tunnelling time also cannot be precisely defined.

\end{document}